\documentclass[aps,prl,twocolumn,showpacs,superscriptaddress,showpacs]{revtex4}

\usepackage{graphicx}
\usepackage{color}

\begin{document}
\title{Stability of the  Bi$_2$Se$_3$(111) topological state: electron-phonon and -defect scattering}
\author{Richard C. Hatch}
\affiliation{Department of Physics and Astronomy, Interdisciplinary Nanoscience Center, Aarhus University, 8000 Aarhus C, Denmark}
\author{Marco Bianchi}
\affiliation{Department of Physics and Astronomy, Interdisciplinary Nanoscience Center, Aarhus University, 8000 Aarhus C, Denmark}
\author{Dandan Guan}
\affiliation{Department of Physics and Astronomy, Interdisciplinary Nanoscience Center, Aarhus University, 8000 Aarhus C, Denmark}
\affiliation{Department of Physics, Zhejiang University, Hangzhou, 310027 China}
\author{Shining Bao}
\affiliation{Department of Physics, Zhejiang University, Hangzhou, 310027 China}
\author{Jianli Mi}
\affiliation{Center for Materials Crystallography, Department of Chemistry, Interdisciplinary Nanoscience Center, Aarhus University, 8000 Aarhus C, Denmark}
\author{Bo Brummerstedt Iversen}
\affiliation{Center for Materials Crystallography, Department of Chemistry, Interdisciplinary Nanoscience Center, Aarhus University, 8000 Aarhus C, Denmark}
\author{Louis Nilsson}
\author{Liv Hornek\ae r}
\affiliation{Department of Physics and Astronomy, Interdisciplinary Nanoscience Center, Aarhus University, 8000 Aarhus C, Denmark}
\author{Philip Hofmann}
\affiliation{Department of Physics and Astronomy, Interdisciplinary Nanoscience Center, Aarhus University, 8000 Aarhus C, Denmark}
\email[]{philip@phys.au.dk}


\date{\today}
\begin{abstract}
The electron dynamics of the topological surface state on   Bi$_2$Se$_3$(111) is investigated by temperature-dependent angle-resolved photoemission. The electron-phonon coupling strength is determined in a spectral region for which only intraband scattering involving the topological surface band is possible. The electron-phonon coupling constant is found to be  $\lambda=0.25(5)$, more than an order of magnitude higher than the corresponding value for intraband scattering in the noble metal surface states. The stability of the topological state with respect to surface irregularities was also tested by introducing a small concentration of surface defects via ion bombardment. It is found that, in contrast to the bulk states, the topological state can no longer be observed in the photoemission spectra and this cannot merely be attributed to surface defect-induced momentum broadening. 
\end{abstract}

\maketitle
Topological insulators are one of of the most intriguing subjects of current condensed matter physics \cite{Hasan:2010,Moore:2010,Zhang:2008}. Despite of their insulating bulk, these materials support metallic edge and surface states with an unconventional spin texture  \cite{Hsieh:2009,Hsieh:2009b}, electron dynamics  \cite{Konig:2007,Roushan:2009} and stability. 
Exploiting these properties is the key to several applications, e. g. in spintronics and quantum computing.  Moreover, several novel physical phenomena are predicted in connection with the topological states  \cite{Fu:2008,Qi:2009b,Linder:2010}.

The stable existence of a gap-closing surface state \cite{Kane:2005b,Kane:2005c,Bernevig:2006} is a property derived from the bulk band structure of a topological insulator, rather than a mere  coincidence. The topological state is also stable in a dynamical sense. A hallmark is the absence of back-scattering near non-magnetic defects. Edge states in the quantum spin Hall effect, a two-dimensional topological insulator, are completely protected from (elastic) scattering \cite{Kane:2005c} whereas the scattering phase space is strongly reduced for surface states on a three dimensional topological insulator \cite{Pascual:2004,Roushan:2009}, preventing localization by weak disorder. 

The stability of the topological state is essential for many phenomena and applications, however, only a few experimental studies have addressed this issue. Experimental measurements using angle resolved photoemission spectroscopy (ARPES) and scanning tunnelling microscopy (STM) have shown that the topological surface states are robust against a small number of adsorbates \cite{Hsieh:2009c,Wray:2010a} and detectable at room temperature \cite{Hsieh:2009c}, but other results question their stability with respect to surface scattering processes \cite{Butch:2010}. Here we determine the electron-phonon ($el-ph$) coupling  strength on the topological insulator Bi$_2$Se$_3$(111) \cite{Zhang:2009,Xia:2009}. In the absence of defects, $el-ph$ scattering can be expected to be the dominant scattering mechanism at finite temperature and it is therefore of exceptional importance for any application. We concentrate on the spectral region in which only the topological state exists and thus only intraband scattering is possible and
we show that while the $el-ph$ coupling constant $\lambda$ is of moderate size, it is surprisingly high compared to well-studied model systems such as the noble metal (111) surface states. We also demonstrate that scattering from a relatively small concentration of surface defects leads to a situation in which the topological surface state can no longer be observed by ARPES. 
  

ARPES experiments were performed on $in$ $situ$-cleaved single crystals of Bi$_2$Se$_3$ at the ASTRID synchrotron radiation facility. Spectra were measured with an angular resolution of 0.13$^{\circ}$ and a combined energy resolution better than 15~meV. All data were taken at a photon energy of $h\nu=16$~eV, except for the photon energy scan in Fig. 3 for which $14\leq h\nu \leq 32$~eV with a step size of $\Delta h \nu = 0.1$ eV. The biggest challenge in this type of study is the time-dependent band bending \cite{Hsieh:2009c,Bianchi:2010b} and the accompanying change of the surface electronic structure. This necessitates a fast measurement. Here the first ARPES measurements were typically taken 10 minutes after cleaving the sample and the entire temperature dependence was concluded after another 5 minutes. Only the data shown in Fig. 3 were taken several hours after cleaving the sample. The STM measurements were performed at room temperature, also on samples cleaved \emph{in situ}. More experimental details can be found in the supplementary online material (SOM) \cite{SOM}. 

\begin{figure}
\begin{center}
\includegraphics[width=\columnwidth]{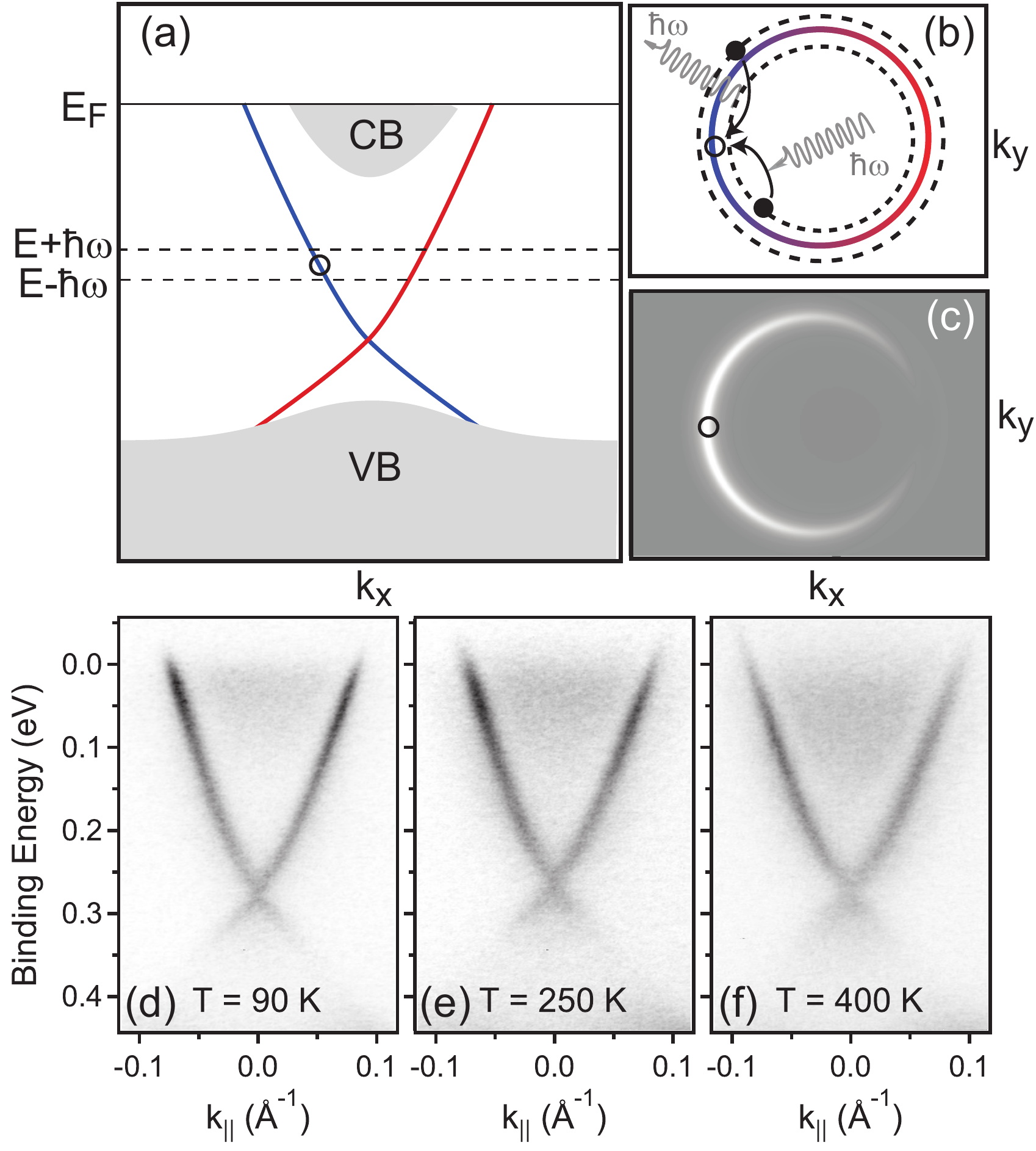}
\caption{(color online). (a) Band structure of Bi$_2$Se$_3$ showing the topological surface state. A photohole of energy $E$ can be filled by electrons which absorb or emit a phonon and originate from an energy window of $\pm \hbar\omega$ where $\hbar\omega$ is the maximum phonon energy in the material. (b) Possible $el-ph$ scattering processes to fill a hole. (c) Probability that an electron with a certain $k_{\parallel}$ will fill the photohole (bright areas having the highest probability). (d)- (f) ARPES spectra taken at different temperatures.} 
\label{fig:1}
\end{center}
\end{figure}

The electronic structure of the Bi$_2$Se$_3$ crystals is shown schematically in Fig. \ref{fig:1}(a). The topological state is situated in the gap between the valence band (VB) and conduction band (CB) and it shows the typical Dirac-cone dispersion. It is challenging to grow intrinsic samples of Bi$_2$Se$_3$ and it is often found, as in our case, that the material is degenerately $n$-doped with the Fermi level in the bottom of the conduction band. This problem can be solved by counter-doping with Ca \cite{Hor:2009} but this has the disadvantage of introducing additional defects and it is therefore not a desirable strategy for the lifetime investigations performed here. 

Fortunately, however, the doping is not a problem for an investigation of the $el-ph$ scattering of the topological state. For this state only the intraband scattering can be studied without involving any bulk states at all. To see this, consider the case of a decaying hole in the topological state in  Fig. \ref{fig:1}(a)-(c). The hole is filled by another electron with a phonon providing energy and momentum. The electrons available for filling thus have to originate from a very limited energy interval of $\hbar \omega$ above (below) the hole for the emission (absorption) of a phonon with this energy. As the maximum phonon energy in the material is quite small (22 meV \cite{Richter:1977}) this energy range is limited. As long as the initial state energy lies well within the bulk band gap only intraband scattering of the topological state is allowed. Note that this is much different from electron-electron scattering where all electrons with a binding energy less than that of the hole, including the ones from the bulk conduction band, contribute to the lifetime. 

The decay of the hole can be expected to be limited further by considerations involving the electron spin. The probability that an available electron fills the hole is greater if its spin orientation is similar to that of the hole state.  The weights for these transitions are represented in a sketch of the $F$ factor (see Ref. \cite{Nechaev:2009}) which is shown in Fig. \ref{fig:1} (c).  Electrons which originate from the bright regions are more likely to fill the hole. 

\begin{figure}
\begin{center}
\includegraphics[width=\columnwidth]{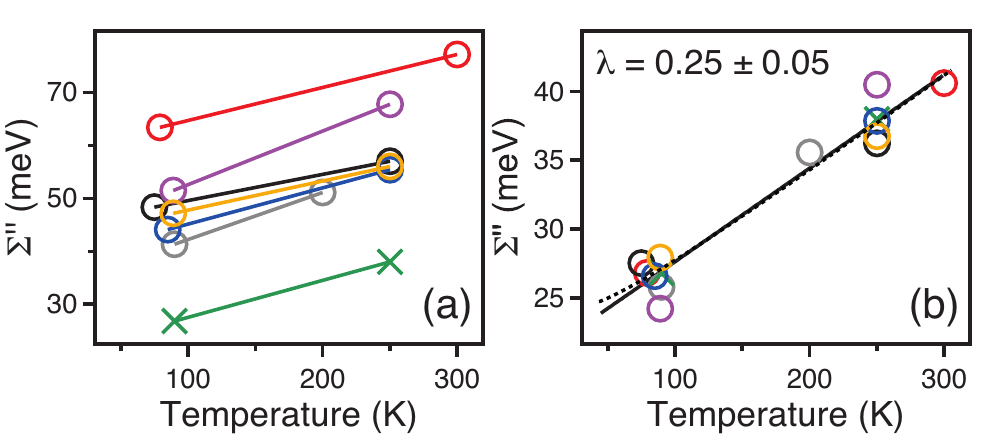}
\caption{(color online). (a) Imaginary part of the self energy ($\Sigma''$) as a function temperature where each set of two points represents a different sample or cleave. The estimated uncertainty for each data point is $\pm$ 3 meV. (b) Data from (a) where the data sets with open circles have been shifted in energy such that a line fit through all the data minimizes $\chi^2$. The dotted line show a calculation for a full Debye model (see text).}
\label{fig:2}
\end{center}
\end{figure}

Temperature-dependent ARPES measurements give straight-forward access to the $el-ph$ coupling strength. Photoemission spectra taken between 90 K and 400~K are shown also in Fig. \ref{fig:1}.   A quantitative analysis of the data was performed by fitting constant energy slices through the spectra (momentum distribution curves) with two Lorentzian lines in order to obtain the position and width of the topological state over an energy range from 50 to 100~meV above the Dirac point, well below the conduction band minimum. The dispersion of the state was obtained from the peak position and fit with a polynomial to obtain a noise-free group velocity as a function of energy. This was combined with the width in order to determine the energy-dependent imaginary part of the self-energy ($\Sigma''$). Not surprisingly, $\Sigma''$ was found to be constant over the small energy range investigated so an average was taken to more precisely determine $\Sigma''$. 

Figure \ref{fig:2} (a) shows the resulting $\Sigma''$ as a function of temperature in sets corresponding to a number of different samples and cleaves. The magnitude of $\Sigma''$ varies greatly depending on the sample and cleave quality which is a result of different and inhomogeneous defect concentrations.  What is also apparent is that $\Sigma''$ has a consistent dependence on temperature.  To see the temperature dependence more clearly, the data sets displayed with open circles in Fig. \ref{fig:2} (a) have been rigidly offset in energy such that a line fit through all the data points minimizes $\chi^2$ (see Fig. \ref{fig:2} (b)).

The $el-ph$ coupling strength can be expressed in terms of the $el-ph$ coupling constant $\lambda$.  
In the present context, $\lambda$ is a spectroscopic parameter defined via the Eliashberg function of the system rather than through  the increase of the effective mass at the Fermi energy. At high temperatures, independent of phonon spectrum, $\lambda$ can be extracted from the temperature-dependent $\Sigma''$ using the relation $\lambda=(\pi k_B)^{-1}d\Sigma''/dT$.  While this is formally valid only at high temperatures, it is expected to hold reasonably well for temperatures comparable to the Debye temperature \cite{McDougall:1995} and thus also in our case ($\Theta_D=182$~K \cite{Shoemake:1969}). A line fit to the data results in $\lambda=0.25\pm0.05$.  The actual value might be slightly smaller due to the effect of the time-dependent band bending \cite{SOM}.

To confirm the justification of the simple linear fit, we also calculate $\Sigma''$ in the full Debye model using $\Theta_D=182$~K and $\lambda=0.26$   \cite{Hofmann:2009b}. The result is shown as a dotted line in Fig. \ref{fig:2}(b) along with the linear fit. The difference is very small and the change of $\lambda$ to get a similar slope at higher temperatures is well within the uncertainties. Note that the real difference is probably even smaller because the Debye temperature at the surface is often only $\approx$70\% of the bulk value \cite{Walfried:1996}. In order to fit the Debye model to the absolute value of $\Sigma''$ reported in Fig. \ref{fig:2} (b), we have to assume a temperature-independent contribution to the self-energy of 20~meV, accounting for both electron-electron and electron-defect scattering. Thus, the $el-ph$ coupling contribution is rather significant, amounting to half of the total self-energy at room temperature. Indeed, its relative significance can be expected to be higher still for intrinsic Bi$_2$Se$_3$ where the electron-electron contribution to the self energy should be much smaller than in our strongly $n$-doped sample.

At first glance, a value of $\lambda=0.25$ appears to imply a moderate $el-ph$ scattering for the topological state. The value is higher than for a good conductor such as the noble metals ($\lambda \approx 0.1$) but much smaller than for a BCS superconductor ($\lambda \approx 1$) \cite{Grimvall:1981}. For the particular system of a topological surface state, however, the value is surprisingly high because the phase-space for the scattering is very small. Only intraband scattering is allowed and the number of available states is further restricted by the requirement for spin conservation outlined above. Rather than to compare to the absolute $\lambda$ value of other systems, one should only consider the intraband contribution for a two-dimensional state. This is well understood for the (111) surface states of the noble metals for which the intraband contribution to $\lambda$ is only $\approx$~0.02  \cite{Eiguren:2003}. Our results thus imply a very significant $el-ph$ scattering strength for the topological state. This is a rather unexpected finding and in strong contrast to an earlier analysis of the topological state's self-energy where the seemingly reasonable assumption of a vanishing intraband scattering was made \cite{Park:2010}.

We briefly address the related issue of scattering by static defects. To this end, 
defects were created  by using a 15 second cycle of sputtering with 0.7~keV Ne$^+$ at a pressure of $3\times10^{-6}$ mbar.  The introduction of a \emph{small concentration} of surface defects due to the sputtering is expected to strongly influence the two-dimensional (2D) states associated with the surface while they should only have a minor effect on the three-dimensional (3D) bulk states. 

\begin{figure*}
\begin{center}
\includegraphics[width=\textwidth]{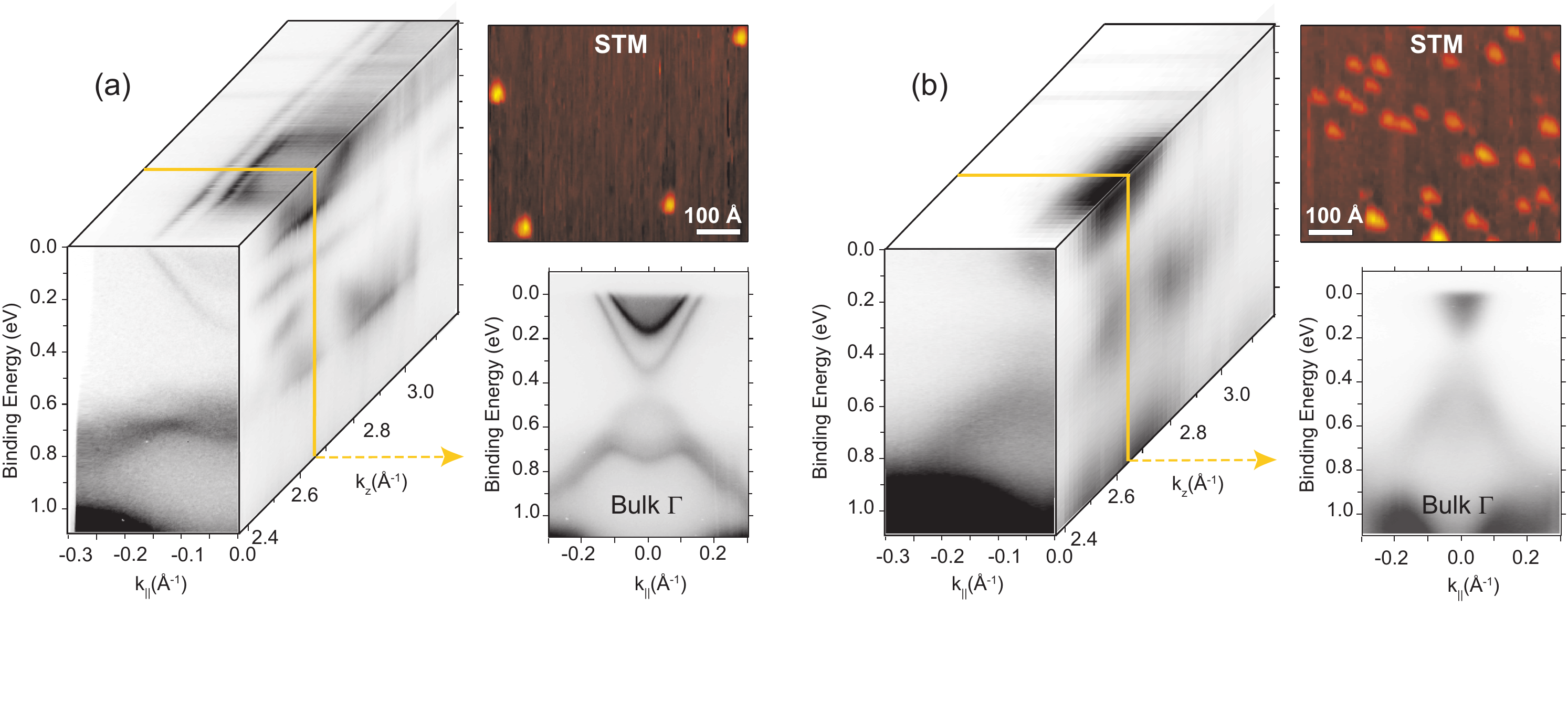}
\caption{(color online). The effect of defect creation on the surface. Differences between (a) a pristine surface and (b) a sputtered surface. Each panel shows a three-dimensional representation of the electronic structure obtained by taking data as a function of photon energy.  Images corresponding to the center of the bulk Brillouin zone are shown separately. The STM images of the surface were measured using a voltage and tunneling current of 1.84~V and 0.18~nA in (a) and 1.68~V and 0.45~nA in (b).}
\label{fig:3}
\end{center}
\end{figure*}

Figure \ref{fig:3} summarizes the photoemission and STM results for the pristine and sputtered surfaces. 2D and 3D states are best distinguished by the absence of $k_z$ dispersion of the former. Therefore results from multiple photon energies are shown and  the $k_z$ dependence of the band structure is calculated assuming a free-electron final state model \cite{Xia:2009,Bianchi:2010b}.  There are three features that do not disperse with respect to $k_z$, thus confirming their 2D character: the topological state, an ``M''-shaped feature at a binding energy of $\approx$~0.75~eV and a surface-localized 2D electron gas formed in the quantum well between vacuum and the downwards-bent bulk conduction band \cite{Bianchi:2010b}. The latter feature is only observed some time after the cleave and thus neither present in the data of Fig. \ref{fig:1} nor in that used to extract the self-energies in Fig. \ref{fig:2}. A comparison of the data before and after sputtering shows a removal of the 2D features whereas the 3D features are merely broadened, most probably due to the non-conservation of $k_{\parallel}$.
 
The disappearance of both the 2DEG and the ``M''-state can easily be attributed to the perturbation of the Bi$_2$Se$_3$ surface.  What is much more difficult to account for is the disappearance of the topological surface state which, due to its topological nature, is expected to be robust against surface defects such as those created by sputtering. The surface defects can act as scattering centers and reduce the mean free path $\lambda_{\textrm{eff}}$ of the surface state electrons, leading to an uncertainty in the crystal momentum on a scale of $\Delta k_{\parallel} = 1/\lambda_{\textrm{eff}}$ \cite{Kevan:1986}.  If the surface defect density is high enough, the disappearance of the topological state could be explained by momentum broadening.  We estimate the average separation between surface defects from STM images taken prior to (after) sputtering to be 425~\AA{} (85 \AA) (Fig. \ref{fig:3}).  If every defect acts as a scattering center we thus obtain a possible crystal momentum broadening of $\Delta k_{\parallel} = 0.002 $~\AA$^{-1}$ and $\Delta k_{\parallel} = 0.012 $~\AA$^{-1}$ for the pristine and sputtered surfaces, respectively.  A momentum broadening of $\Delta k_{\parallel} = 0.012 $~\AA$^{-1}$ would be far too small to explain the extinction of the topological surface states. Hence, it appears that the topological states are strongly perturbed, even by a small concentration of defects, consistent with recent transport measurements \cite{Butch:2010}. Note that this observation is not necessarily inconsistent with previous findings of stability after a long time in vacuum \cite{Park:2010} or even exposure to air \cite{Analytis:2010} if one assumes the surface of Bi$_2$Se$_3$ to be chemically inert. Note also that the failure to observe the state in ARPES does not imply that it does not exist anymore either. Further research, including more local tools such as spectroscopic STM are needed to investigate this issue.

In conclusion, we find that the $el-ph$ coupling for the topological state of Bi$_2$Se$_3$ is surprisingly strong in view of the fact that only intraband scattering can contribute to the decay of the photohole. On the other hand, the state still dominates the spectral function at room temperature such that the $el-ph$ coupling is not expected to preclude room temperature applications. The states' apparent sensitivity towards a  moderate amount of defects is a more significant consideration for practical applications. Despite the relatively inert surfaces of the topological insulators, it appears that a useful practical device will exploit states located at an interface between a topological and a trivial insulator rather than on the interface between a topological insulator and vacuum. Such a configuration would have the added advantages that it would stabilize the fragile band bending near the interface and permit a gating of the near surface electronic structure.

We acknowledge stimulating discussions with I.~A.~Nechaev, S.~V.~Eremeev and E.~V.~Chulkov as well as financial support by the Lundbeck foundation, the Danish Council for Independent Research (Natural Sciences) and the Danish National Research Foundation.



\end{document}